\documentclass[multphys,vecphys]{svmult}
\usepackage[latin2]{inputenc}
\newcommand{\be}{\begin{equation}} 
\newcommand{\ee}{\end{equation}}
\usepackage{amsfonts}
\usepackage{amssymb}

\usepackage{makeidx} 
\usepackage{graphicx} 
                             
\usepackage{multicol} 
\usepackage[bottom]{footmisc}

\makeindex             

\begin{document}

\title*{Influence of information flow in the formation of economic cycles}
\author{J. Mi\'skiewicz\inst{1}\and
M. Ausloos\inst{2}}
\institute{Institute of Theoretical Physics, University of Wroc\l{}aw, pl M.
Borna 9, 50-204 Wroc\l{}aw, Poland 
\texttt{jamis@ift.uni.wroc.pl}
\and SUPRATECS, B5, University
of Li$\grave e$ge, B-4000 Li$\grave e$ge, Euroland
\texttt{marcel.ausloos@ulg.ac.be}}

\maketitle

\section{Introduction}

Despite the fact that the Verhulst's  idea \cite{verhulst} of regulated biological
populations is 200 years old, it is still very useful since it allows to investigate features
of various systems. Here an eight order logistic map is applied in modelling 
the influence of information flow delay onto the behaviour of an economic system.

The delay of information flow is an internal feature of all economic systems,
because continuous monitoring of such systems both on macro and
microeconomy scales is either extremely difficult or even impossible. The more so since
the data is not easily available nor even reliable as it could be in physics
laboratories. Macroeconomy parameters such as Gross Domestic Product, Gross National Product,
inflation, demographic data etc. are announced in well defined time intervals
(monthly, quarterly or annually). The same situation is observed in the case of
various companies. They announce their financial statements about their economic results 
at specific dates and for given time intervals -- according to internal or external rules (usually
according to law regulations). Sometimes some ''warning'' is issued. 
However the tendency is that intervals between announcements are rather long, e.g. the value of a dividend is announced annually or at various trimester ends. It seems obvious that only very small companies are able to perform continuous monitoring.  But even then, the process of collecting information from a significant (on a macroscopic
scale) number of such companies inhibits or makes it impossible to perform continuous
 monitoring. In view of the data collecting procedure it is clear that every
economic decision is based on some information describing a past situation. It
is also important to notice that the time delays between information gathering,
 decision taking, policy implementation, and subsequent data gathering are
 not regularly spaced, nor is a fortiori a continuous variable, as that was
considered in \cite{time_delay}; indeed the information about the system is updated 
at the end of discrete time intervals.

Therefore econophysics-like modelling of such features
encounters some difficulty, surely at the testing level.
Recently a microscopic-like approach has been presented, through a model \cite{ACP1,ACP2,ACP3} including
some measure of a company fitness with respect to an external field, 
and a birth-death evolution, according to some business plan, and the local company close range environment. The information flow was however considered to occur instantaneously. 
 
In order to investigate the discrete information flow time delay 
and its effect, a model, hereby called the ACP model
\cite{ACP1,ACP2,ACP3}, has been modified by splitting the information about the
system into two parameters. One is monitored continuously (is updated at every
iteration step) and is known to the system itself; the second, -- like  official
statements of the system,  is announced at the end of discrete time intervals and is used by
companies for calculating their strategies. Therefore the strategy of a company depends
on the delay time information and the information itself. As it is shown in Sect. \ref{results} the length of
the time delay $(t_d)$ influences quite strongly and in a nontrivial way
the behaviour of the overall system. 

Detailed description of the ACP model is given in Sect. \ref{acp} and the
properties of the system as a function of time delay and initial concentration
are investigated (Sect. \ref{results}) in the case of short, medium and long time
delays.

\section{ACP model}
\label{acp}
\index{ACP model}
For the sake of clarity the basic ingredients of the ACP model are recalled
here below.   
 The main problem was to simulate the behaviour of economic systems in spatio-temporally 
 changing
environmental conditions, e. g. political changes and destruction of economy
barriers. The model was set in the form of a Monte Carlo simulation. Notice that the ACP model
\cite{ACP1,ACP2,ACP3} contains among its variants an adaptation of the Bak -- Sneppen model
\index{Bak -- Sneppen model} and was built in order to answer economy
questions\footnote{Let us recall that the Bak- Sneppen model was originally built
in order to investigate the coevolution of populations \cite{back_sneppen}}. The model consists of 
\begin{enumerate}
\item { \bf space} -- a square symmetry lattice,
\item {\bf companies}, which are initially randomly placed on the lattice, in an
\item  { \bf environment}   \index{environment} characterised by a real field $F \in [0,1]$ and a selection
pressure $ sel $,
\index{selection pressure}
\item each company ($ i $) is characterised by one real parameter $f_i \in [0,1]$
 (so called its {\bf fitness}).
\index{fitness}
\end{enumerate} 
The following set of actions was allowed to companies:
\begin{enumerate}
\item companies survive with the probability 
\be
p_i = \exp (-sel |f_i -F | )
\ee
\item companies may move on the lattice horizontally or vertically,
one step at a time, if  some space is available
in the von Neuman neighbourhood. 
\index{von Neuman neighbourhood}
\item if companies meet they may
\begin{enumerate}
\item either merge with a probability $b$,
\item or create a new company with the probability $1-b$.
\end{enumerate} 
\end{enumerate} 
\index{mean field approximation}
The ACP model may be described in a mean field approximation
\cite{ACP_AM, ACP_AM1} by introducing the
\index{distribution function} distribution function of companies $ N(t,f) $,
which describes the number of companies having a given fitness $ f $ at time $ t $. The
system is then additionally characterised by the concentration of companies $ c(t) $.

The present report of our investigations is restricted to the case of the best
adapted companies $ (f=F) $, so that the selection pressure has no influence on the
survival of companies. So the only factor which could alter the number of
companies is the strategy, i.e. the decision to merge or create a new entity.
The ideas behind the mean field approximation \cite{ACP_AM, ACP_AM1} is applied here and
developed by introducing a strategy depending on the system state and the
discrete time of the official announcement about the state of the system. 

The introduction of the strategy depending on the state of the system reflects the
idea of Verhulst \cite{verhulst}, when replacing the constant Malthus grow rate by the
function $1-x$, which introduced a limit for the system to grow. In the present
investigation  it is assumed that the strategy should depend on the state of the
system. Moreover the company board takes its decision knowing informations announced about its
environment. The generation of new entities is more likely in the case of a low
 concentration  of companies than when this concentration is high. The merging parameter
  describes the reversed dependency, i.e.  merging is more likely to occur in the case of a
high density of companies than if the density is low. The simplest function 
which fulfils this condition is $1-c$, the same as in Verhulst original work
\cite{verhulst}.

The additional ingredients to the ACP model are thus
\begin{enumerate}
\item the merging parameter $b$ is replaced by a strategy $(1-c)$, \index{merging
parameter}
\item the companies know the value of the concentration $c$ according to official statements
announced after the time delay $t_d$.
\end{enumerate} 
The evolution equation of the system with companies, using the state dependent
strategy is:
\index{state dependent strategy}
\be
\label{evol}
c_t = c_{t-1} + \frac{1}{2} c_{t-1}(1-c^8_{t-1})(1-(1-c_{t-1})^8) (2
ST(c(g(t)))-1),
\ee
where
$ST(c) = 1-c$, $g(t) = k [\frac{t}{k}]$ and $[ \, ]$  denotes the procedure of
taking a natural number not larger than the one given in the brackets.
The time is measured in  iteration steps $IS$.

\section{Results}
\label{results}

Numerical methods were used in order to investigate properties of the system.
Because the coevolution equation (\ref{evol}) is given as an iteration equation
the time is discrete and counted in iteration steps (IS).  The following features
of the system were examined:
\begin{enumerate}
\item the { \bf coevolution} of $c(t)$  as a function of the initial concentration,
\item the {\bf stability time} defined as the time required to achieve a unique stable solution; 
  because of numerical reasons the criterium applied here is 
$| c_{n+1} - c_n|< 10^{-10} $,
\index{stability time}
\item the {\bf crash time} -- $t_c$, such that $c_{t_c} < 0 $ (it is understood as a time
when all companies are wiped out from the system),
\index{crash time}
\item the {\bf stability intervals} -- the intervals of initial values for which the
evolution of the system is longer than a given time $t_s$
\index{stability intervals}
\item the complex Lyapunov exponent
 \be
 \label{comp_lap}
 \lambda = \lim_{N \rightarrow \infty } \frac{1}{N} \sum_{n=1}^N \log_2
(\frac{dx_{n+1}}{dx_n}).
 \ee
\end{enumerate} 
\index{complex Lyapunov exponent}
The Lyapunov exponent calculated in its complex form 
(\ref{comp_lap}) gives also some information about the oscillations of the system. Using
the properties of logarithm:
\be
\label{log_prop}
 a < 0 \Rightarrow \log (a)=\log(-1 \cdot |a|) = \log(-1) + \log(|a|) .
 \ee
the imaginary part of $ \log_2 (\frac{dx_{n+1}}{dx_n}) $ gives some information on
whether the distances between consecutive iterations are monotonic.

The numerical iterations were performed for the initial concentration in the interval $ c_0 \in (0,1)$, at consecutive values distant of 0.02. Therefore 500 histories of evolution were investigated.

There are possible to observe three types of coevolution -- a unique, a periodic and a chaotic solution. In the case of unique solution the system may approach this solution in the form "damped" coevolution or "damped oscillation". The damped coevolution is if $\forall t > 0 \;\; c(\infty) - c(t) > 0 $ or $\forall t > 0 \;\; c(\infty) - c(t) < 0 $ and $|c(t-1)-c(t)| \geq |c(t)-c(t+1)|$, where $c(\infty)$ is the asymptotic state of the system. This means that the distance between concentration and asymptotic concentration is decreasing in every iteration step and either the concentration is smaller or bigger than the asymptotic concentration. The damped oscillations are observed if  $|c(t-1)-c(t)| \geq |c(t)-c(t+1)|$ and $ \exists t_0 $ such that $\forall t> t_0 \; c(t) > c(\infty) $ and $c(t+1) < c(\infty)$, this means the distance between consecutive concentrations of companies is decreasing. In the case of a periodic solution for $t>t_0$ there exists a n-tuple of concentrations which is repeated for $t>t_0$, where $t_0$ is the time required by the system to reach the stable or periodic solution. The length of the n-tuple is defined as the period of oscillations.
The system is chaotic if the real part of the Lyapunov exponent $Re(\lambda) > 0$.

\begin{figure}
\centering
\includegraphics[scale=0.45, angle=-90]{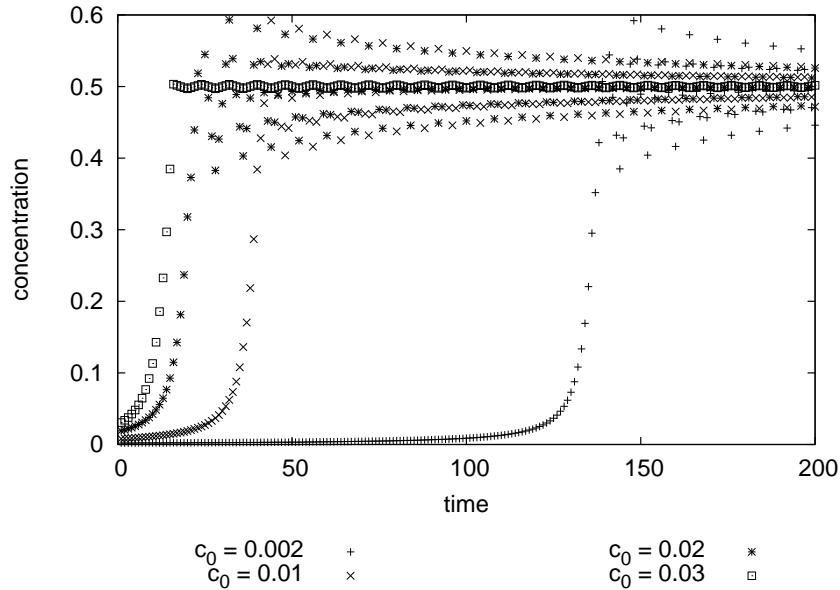}
\caption{Evolution of the system for given initial concentrations;  \newline
delay time $t_d =4 \, IS $}
\label{fig_ver_hist_4}   
\end{figure}

The coevolution of the system is presented either as a function of time (Fig. \ref{fig_ver_hist_2}, \ref{fig_ver_hist_3}, \ref{fig_ver_hist_4}, \ref{fig_ver_hist_5}, \ref{fig:hist12}, \ref{fig:hist15}, where the coevolution is plotted for chosen initial concentrations) or as a function of initial concentration (Fig. \ref{fig_ver_2}, \ref{fig_ver_3}, \ref{fig_ver_4}, \ref{fig_ver_5}, \ref{fig:12}, \ref{fig:15}, where the coevolution of the system is plotted in one vertical line so the plot is a set of coevolutions for 500 different initial concentrations. 

\subsection{Stability window}

The short time delay ($t_d$) is defined as a $ 2 \, IS \leq t_d \leq 4 \, IS)$.
In this case the system evolves to the unique stable solution $c=0.5$. Within
this time delay the Lyapunov exponent is equal to zero; no chaotic behaviour is
seen. 

\subsection*{$t_d = 2 \, IS$}

\begin{figure}
\centering
\includegraphics[scale=0.45, angle=-90]{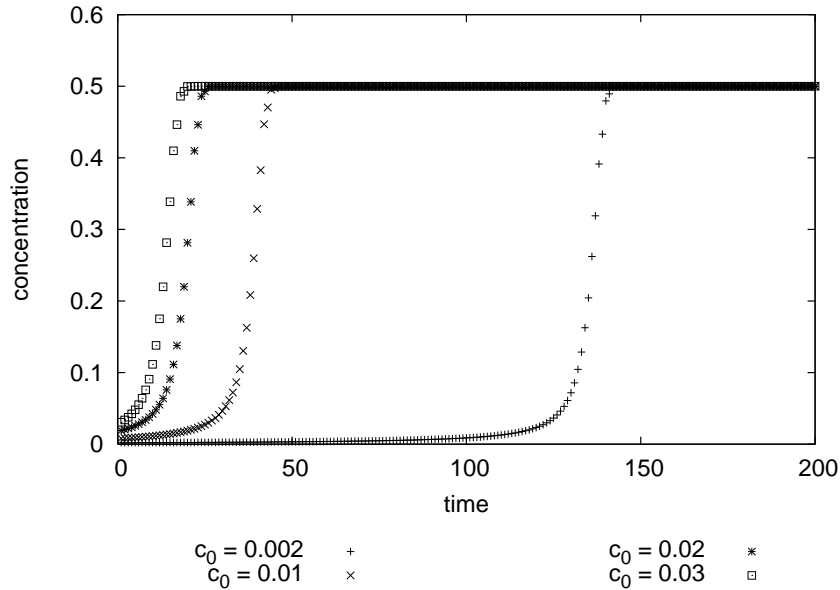}
\caption{Coevolution of the system for chosen initial concentrations;  \newline
delay time $t_d =2 \, IS $}
\label{fig_ver_hist_2}   
\end{figure}

\begin{figure}
\centering
\includegraphics[scale=0.45, angle=-90]{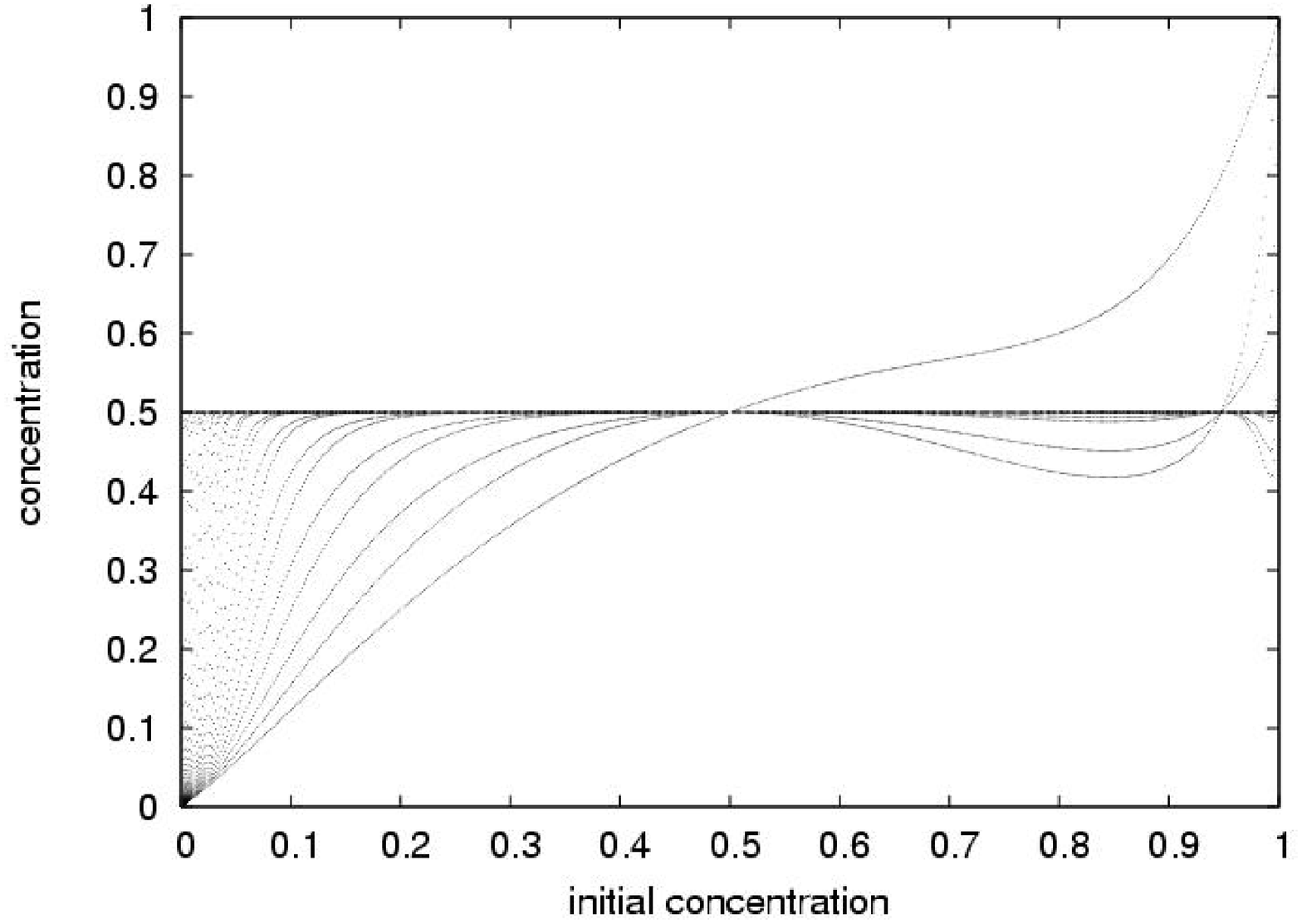}
\caption{Coevolution of the system as a function of initial concentration. The coevolution of a system is represented by a vertical series of dots;  delay time $t_d =2 \, IS $}
\label{fig_ver_2}   
\end{figure}

\begin{figure}
\centering
\includegraphics[scale=0.45, angle=-90]{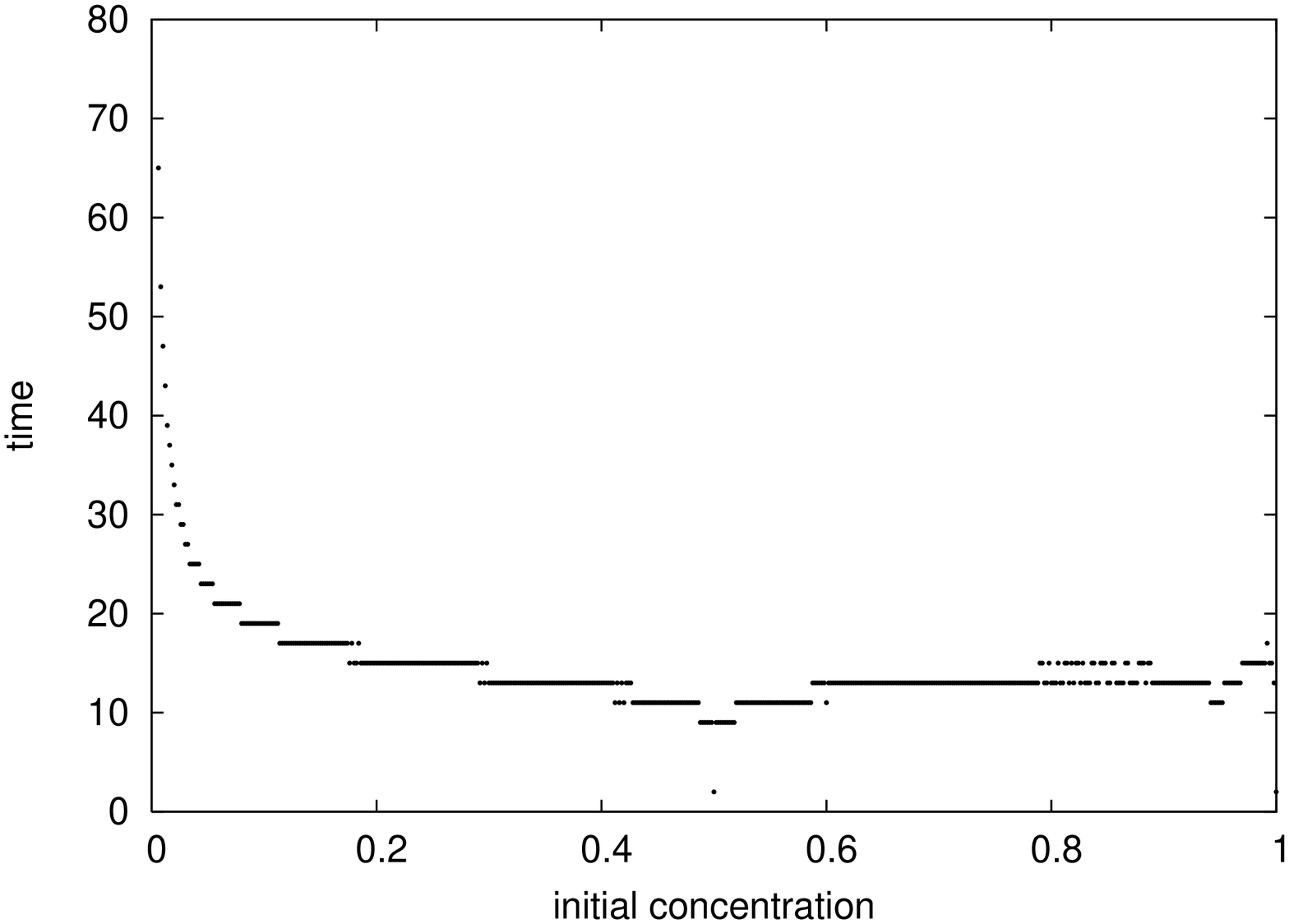}
\caption{The time required for the system to achieve a stable concentration as a function of initial concentration;  delay time $t_d =2 \, IS $}
\label{fig_ver_stab_2}   
\end{figure}

The time delay $t_d=2 \, IS$ means that the information about the system is
updated every two iteration steps. The evolution of the system is presented in
Fig. \ref{fig_ver_2} and is plotted as a function of initial iteration. For every
500 initial concentrations $103$ iteration steps have been used. The
history examples are presented in Fig. \ref{fig_ver_hist_2} as a function of
concentration in time. In the case of the shortest time delay considered here the
system has a unique solution $c=0.5$. The stability time as a function of initial
concentration is shown in Fig. \ref{fig_ver_stab_2}. 
For a very low initial concentration $ 0 < c_0  \lesssim  0.01 $ a long time $(t_s
\geq 47 \, IS)$ is needed in order to achieve the stable state. It is also illustrated in
Fig. \ref{fig_ver_hist_2}, where in the case of low initial concentrations $c_0 =
0.002$ the stability time is quite long time (about 100 IS). However except for very
small initial concentrations ($c_0 > 0.1$) the stability time  is short $t_s \in
(10 \, IS, 20 \, IS)$.

\subsection*{$t_d = 3 \, IS $}

\begin{figure}
\centering
\includegraphics[scale=0.45, angle=-90]{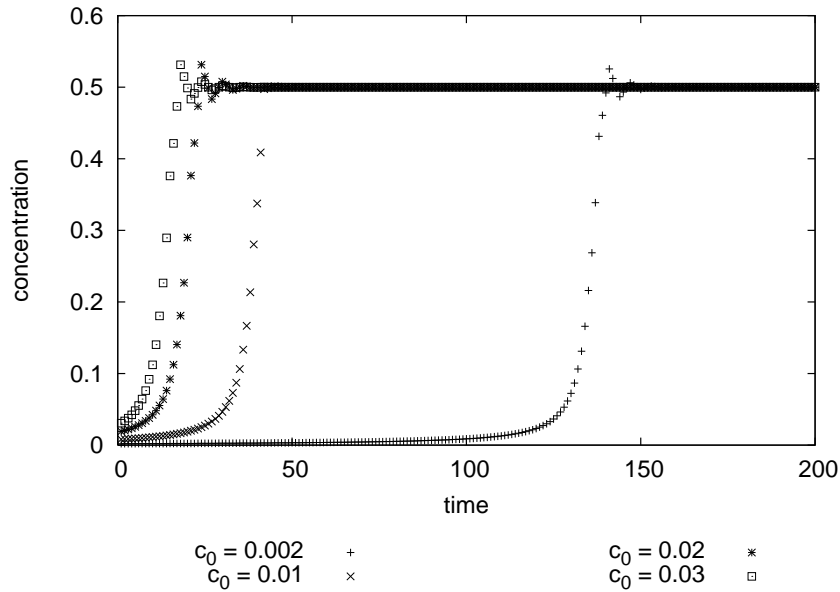}
\caption{Coevolution of the system for given initial concentrations; \newline
 delay time $t_d =3 \, IS $}
\label{fig_ver_hist_3}   
\end{figure}

\begin{figure}
\centering
\includegraphics[scale=0.45, angle=-90]{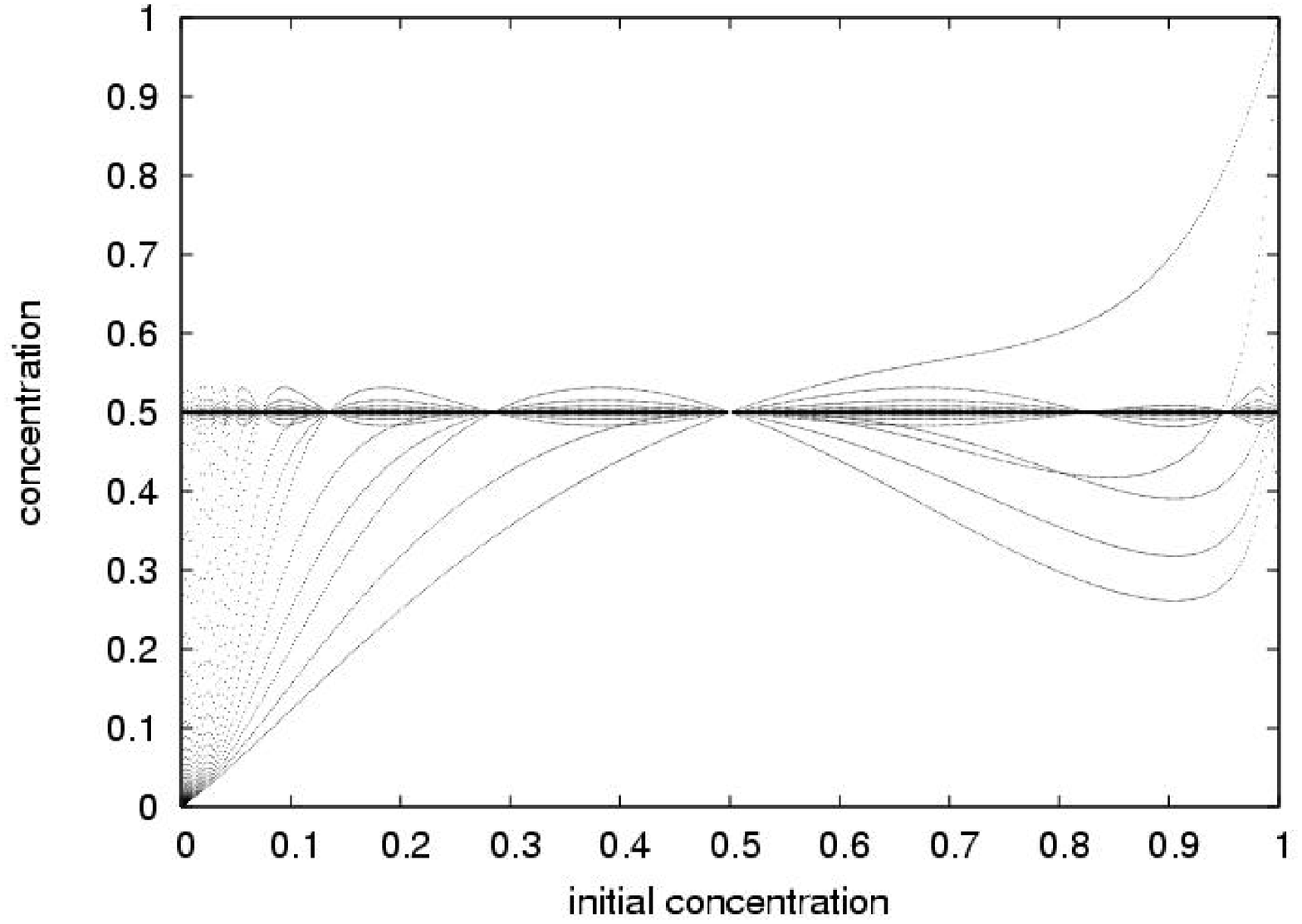}
\caption{Coevolution of the system as a function of initial concentration. The coevolution of a system is represented by a vertical series of dots;   delay time $t_d =3 \, IS$}
\label{fig_ver_3}   
\end{figure}

\begin{figure}
\centering
\includegraphics[scale=0.45, angle=-90]{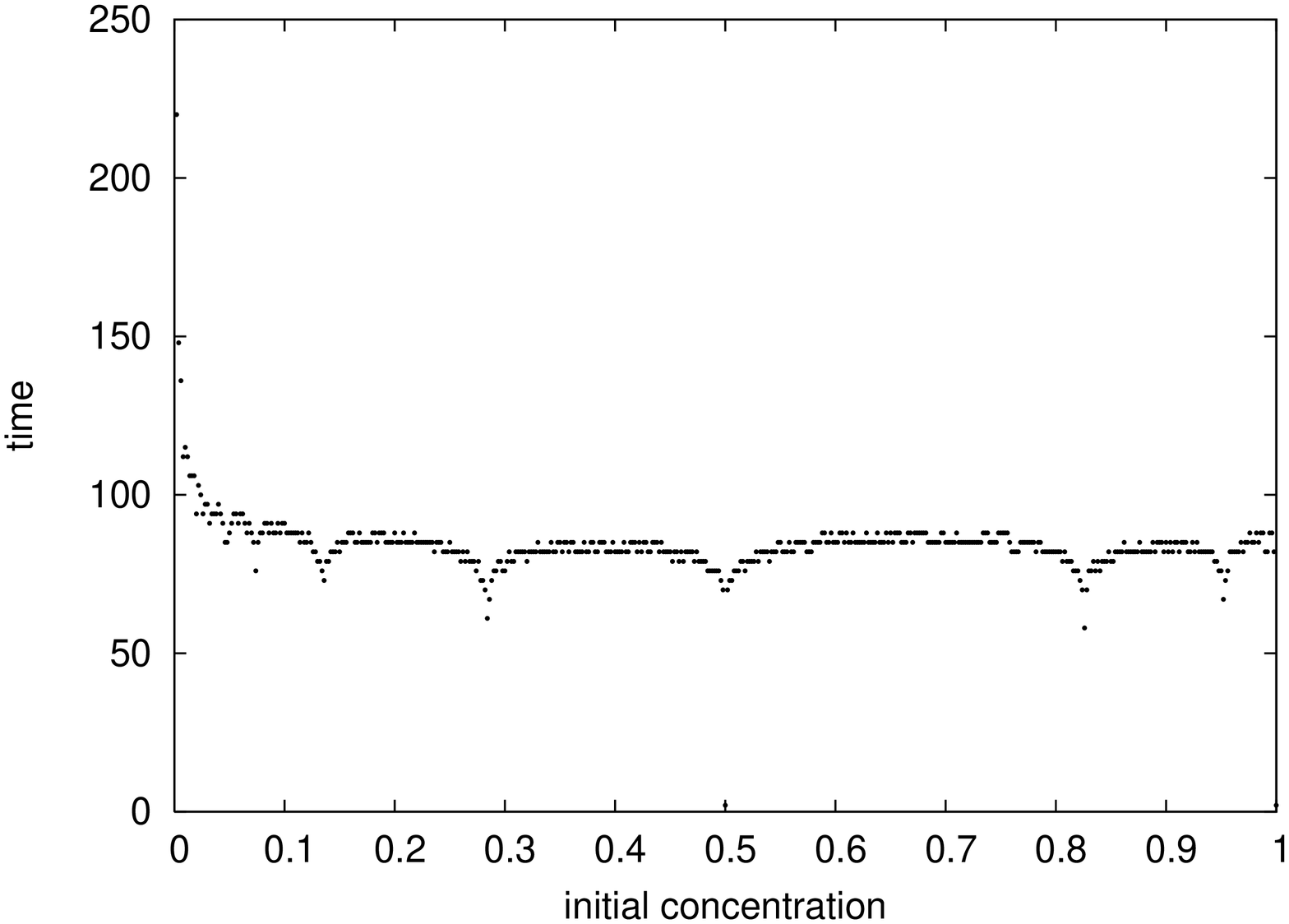}
\caption{The time required for the system to achieve stable concentrations as a function of its initial concentration; delay time $t_d =3 \, IS $}
\label{fig_ver_stab_3}   
\end{figure}

Extending the time delay by one, up to three iteration steps, induces important
changes in the system. In the evolution of the system, damped oscillations become
observable, e.g. for $c_0 = 0.002$  damped oscillations are observed for $t \in (
140 \, IS, 155 \, IS ) $  (Fig. \ref{fig_ver_3} and Fig.  \ref{fig_ver_hist_3}).
The maximum time required for the system to achieve a stable state extends to $t_s
\geq 220 \, IS$ as compared with $t_s \geq 47 \, IS$ for $t_d =2 IS$. For most 
initial concentrations ( $c_0 > 0.05$)  the stability time is in the interval $t_s
\in ( 70 \, IS , 100 \, IS )$. Therefore the system requires a longer time to
achieve a stable state. However there are some "stability points" for which the
system achieves a stable state markedly faster. These can be found on Fig.
\ref{fig_ver_stab_3}; these points are: $c_0 = 0.074, \; t_s = 76 \, IS$; $c_0
= 0.136, \; t_s = 73 \, IS$; $c_0 = 0.284, \; t_s = 61 \, IS $; $c_0 = 0.5, \;
t_s = 1 \, IS$; $c_0 = 0.826, \; t_s = 58 \, IS$; $c_0 = 0.952, \; t_s = 67 \, IS$.

Comparing the results obtained in the case of $t_d=2IS$ and $t_d=3IS$ it can  be noticed that the stability times is significantly extended and new features become visible (damped oscillations). Therefore we can conclude that  the system is very sensitive to the flow of information and extension by only one IS step of the time delay changes the behaviour of the system quite significantly.

\subsection*{$t_d = 4 \, IS$}

\begin{figure}
\centering
\includegraphics[scale=0.45, angle=-90]{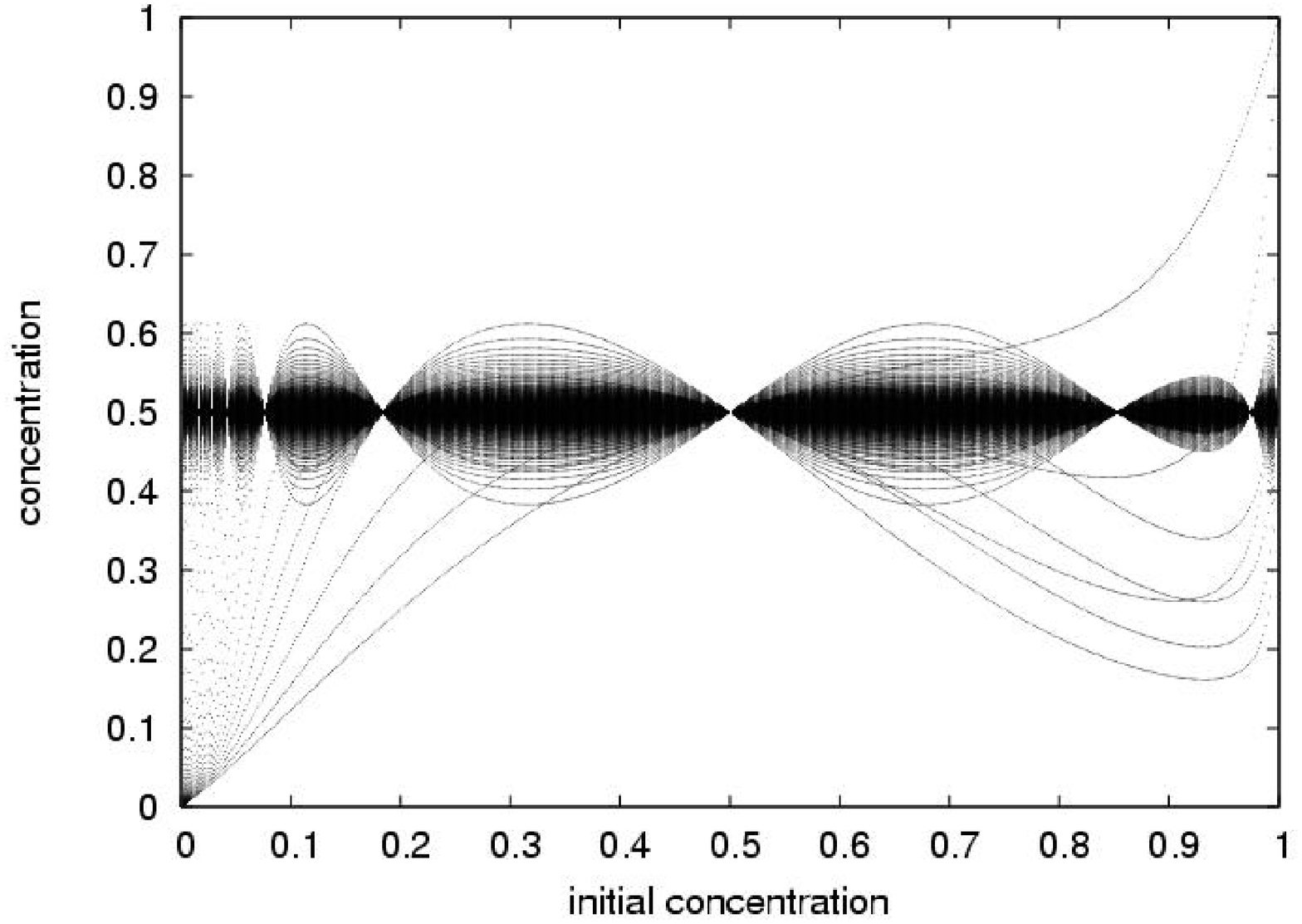}
\caption{Coevolution of the system as a function of initial concentration. The coevolution of a system is represented by a vertical series of dots;  delay time $t_d =4 \, IS $}
\label{fig_ver_4}   
\end{figure}

\begin{figure}
\centering
\includegraphics[scale=0.45, angle=-90]{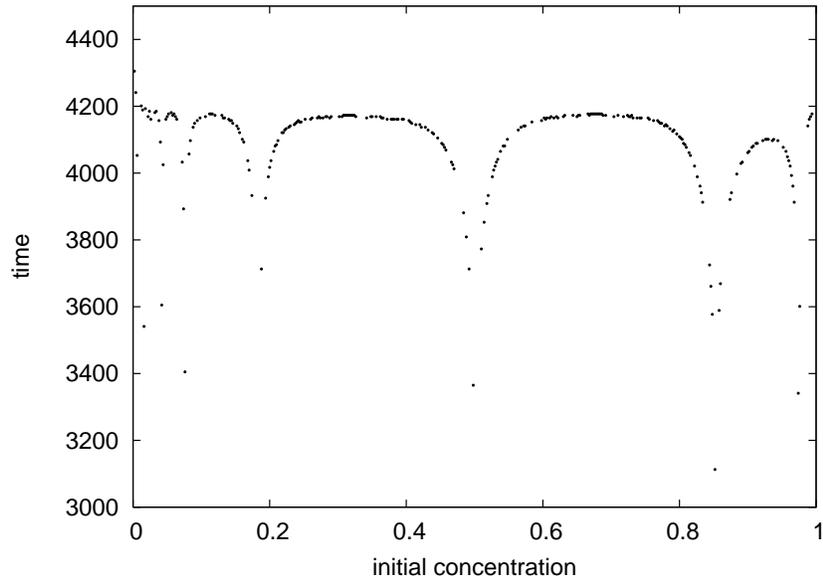}
\caption{The time required for the system to achieve stable concentrations;
\newline  delay time $t_d =4 \, IS $}
\label{fig_ver_stab_4}   
\end{figure}

For a time delay $t_d = 4 \, IS$, the lately seen features
(damped oscillations)  are also present as it can be observed on both figures showing
the coevolution for the considered initial concentrations and for chosen histories
presenting explicitly time evolution of the system -- Fig.\ref{fig_ver_3} and 
Fig.\ref{fig_ver_4} respectively. It is worth noticing that the damping of
oscillations is much weaker than in the case $t_d = 4$ (compare
Fig.\ref{fig_ver_3} and  Fig.\ref{fig_ver_4}).  The oscillation amplitude is
decreasing significantly more slowly for the case $t_d=4 \, IS$ than for $t_d=3 \,
IS$. However in all considered cases $t_d=2 \, IS$, $t_d=3 \, IS$, $t_d=4 \, IS$,
the system has one stable solution, but the stability time depends on the delay
time; it is the longest in the case  $t_d=4 \, IS$ ($ 3200 \, IS \leq t_s \leq 4200 \, IS $). The time required for the system to achieve stable state is presented in Fig. \ref{fig_ver_stab_4}. As in the previous case $t_d = 3 \, IS$ there are initial concentrations for which the system reaches the stable state significantly quicker, e.g. $c_0=0.23$. 

\subsection{Medium time delay}

\subsection*{$t_d = 5 \, IS$ and $t_d = 6 \, IS$}

\begin{figure}
\centering
\includegraphics[scale=0.45, angle=-90]{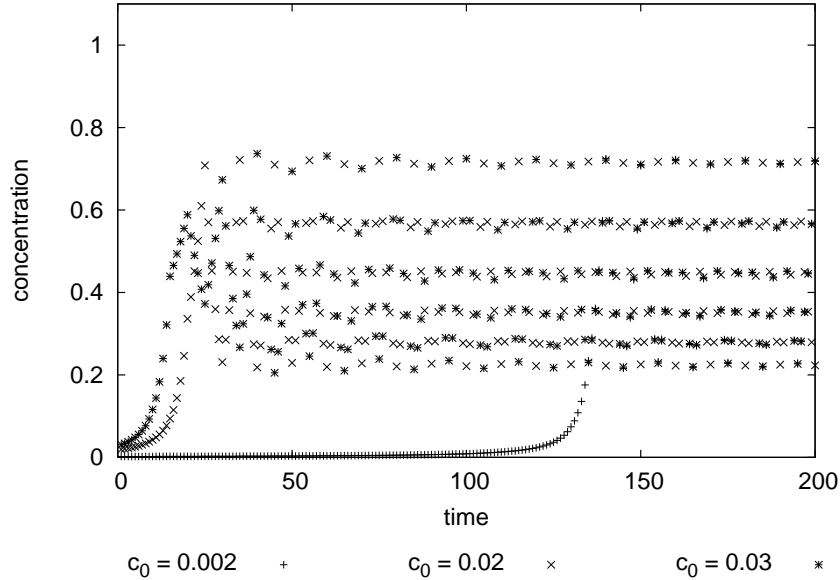}
\caption{Evolution of the system for chosen initial concentrations;  \newline
delay time $t_d =5 \, IS $}
\label{fig_ver_hist_5}   
\end{figure}

\begin{figure}
\centering
\includegraphics[scale=0.45, angle=-90]{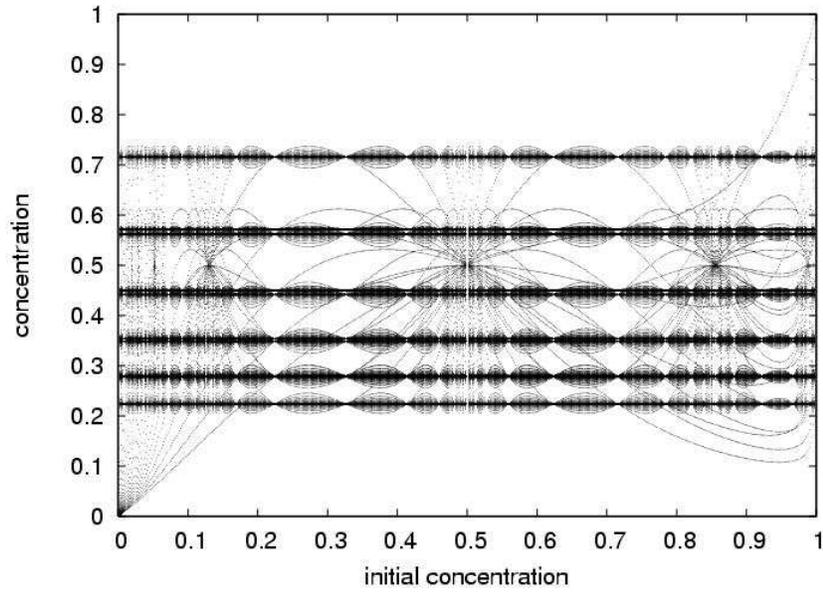}
\caption{Coevolution of the system as a function of initial concentration. The coevolution of a system is represented by a vertical series of dots; delay time  $t_d =5 \, IS $}
\label{fig_ver_5}   
\end{figure}

The five iteration step delay time $(t_d = 5 \, IS)$ is very interesting,
because this is the shortest time for which cycles of concentration can be
observed. For this time delay the system has several solutions. Despite the fact that
the real part of the Lyapunov exponent is still equal to zero, its imaginary part is not. According to Eq. (\ref{log_prop}) this shows that the system
has a periodic solution. These solutions can be seen in Fig. \ref{fig_ver_5} and
Fig. \ref{fig_ver_hist_5}. In the case of Fig. \ref{fig_ver_5} the evolution is
shown as a function of its initial concentration, whereas Fig. 
\ref{fig_ver_hist_5} presents the system evolution as a function of time for
chosen initial concentrations. 

\begin{figure}
	\centering
	\includegraphics[angle=-90,scale = 0.45]{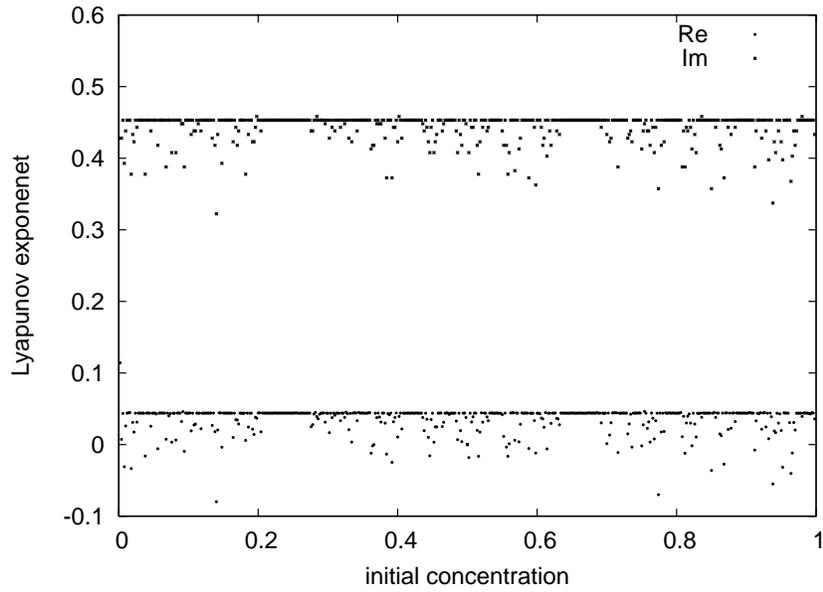}
	\caption{Lyapunov exponent for $t_d = 6 \, IS$}
	\label{fig:lap6}
\end{figure}

Oscillating solutions can be also found in the case $ t_d =6 \, IS$;  the
imaginary part of the Lyapunov exponent, as in the previous case ($t_d=5 \, IS$) is
negative (Fig. \ref{fig:lap6}).

\subsection{Long time delay}

\subsection*{$t_d \geq 7 \, IS$}

\begin{figure}
	\centering
	\includegraphics[angle=-90,scale = 0.45]{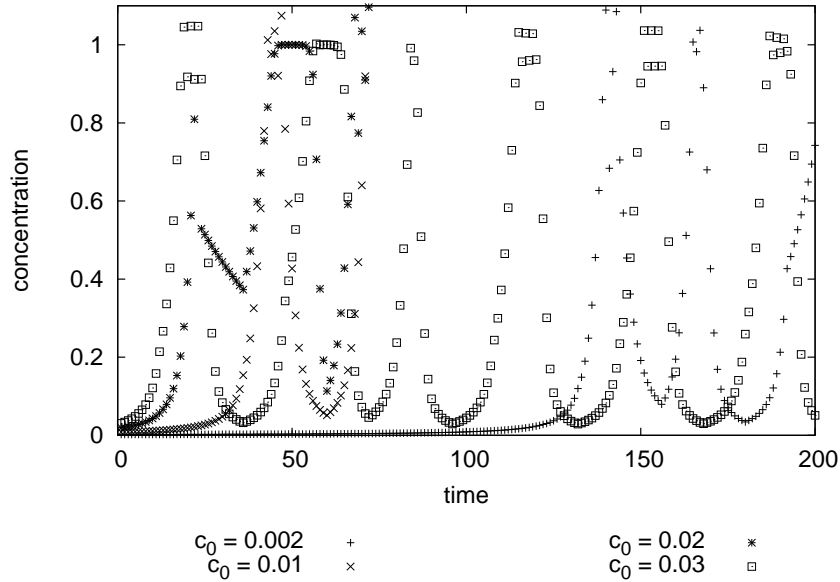}
	\caption{Coevolution of the system as a function of time for chosen
initial concentrations; $t_d=12 \, IS$}
	\label{fig:hist12}
\end{figure}

\begin{figure}
	\centering
	\includegraphics[angle=-90,scale = 0.45]{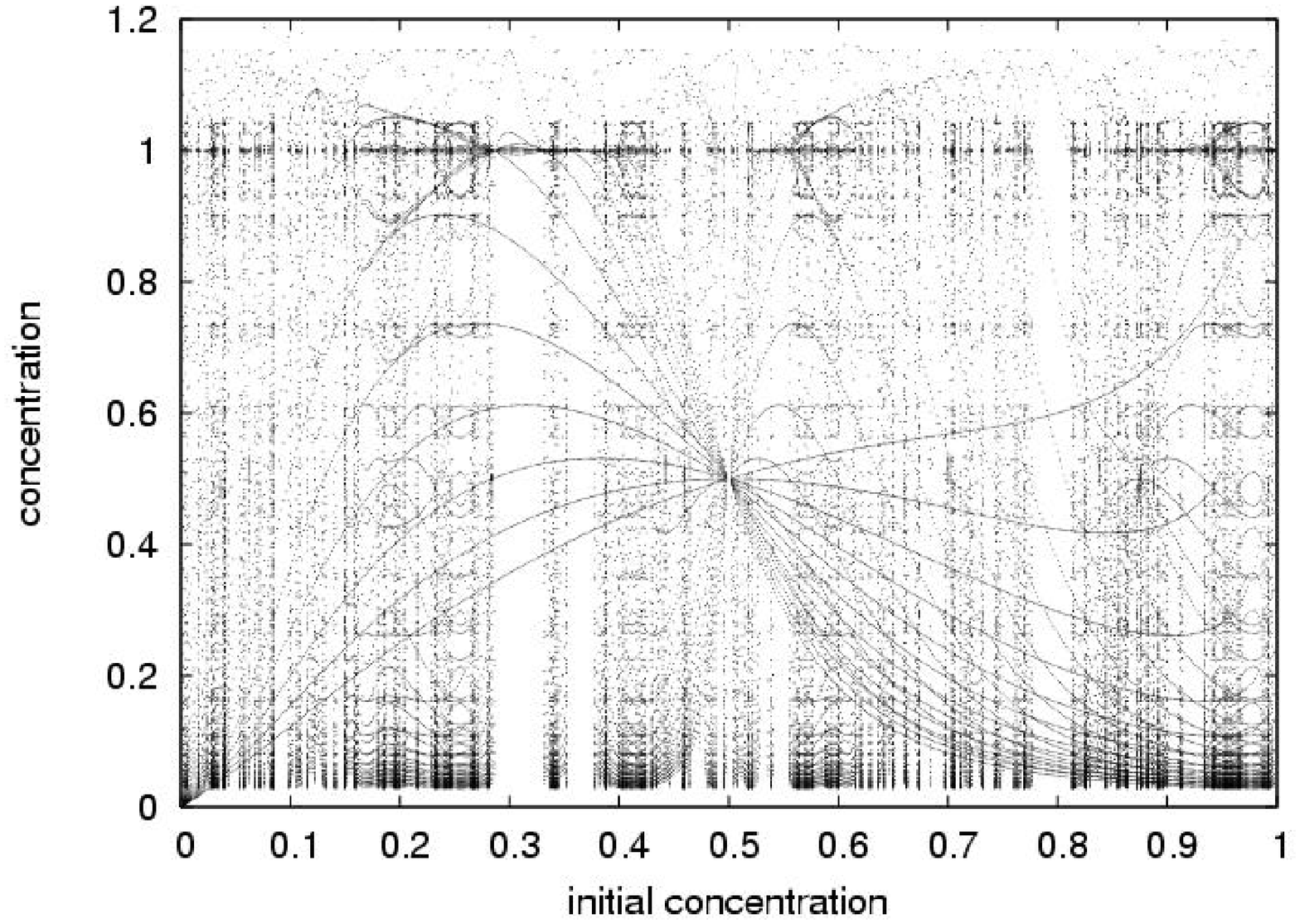}
	\caption{Coevolution of the system as a function of its initial concentration. The coevolution of a system is represented by a vertical series of dots; $t_d = 12\, IS$}
	\label{fig:12}
\end{figure}

\begin{figure}
	\centering
	\includegraphics[angle=-90,scale = 0.45]{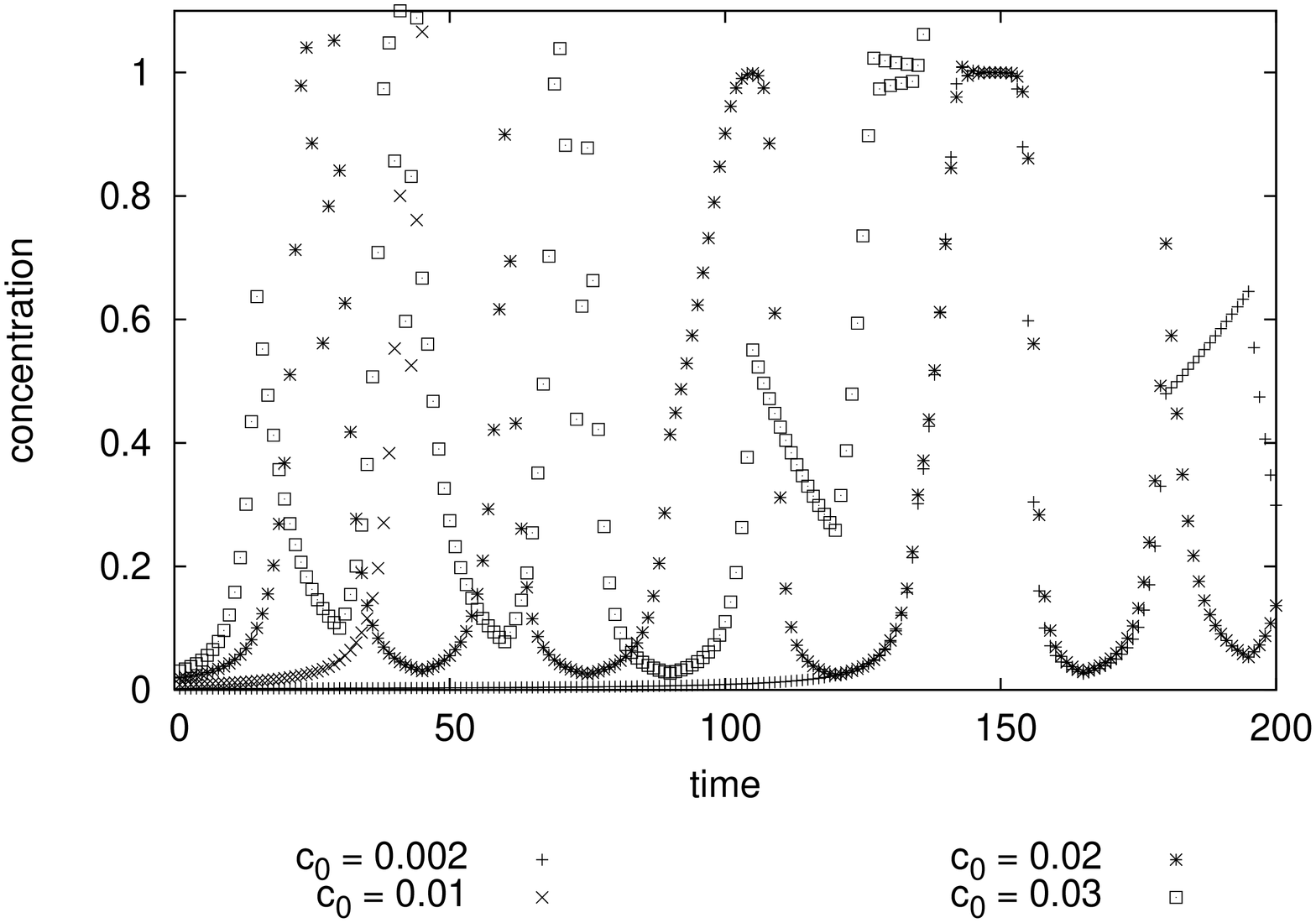}
	\caption{Coevolution of the system as a function of time for chosen
initial concentrations; $t_d=15 \, IS$}
	\label{fig:hist15}
\end{figure}

\begin{figure}
	\centering
	\includegraphics[angle=-90,scale = 0.45]{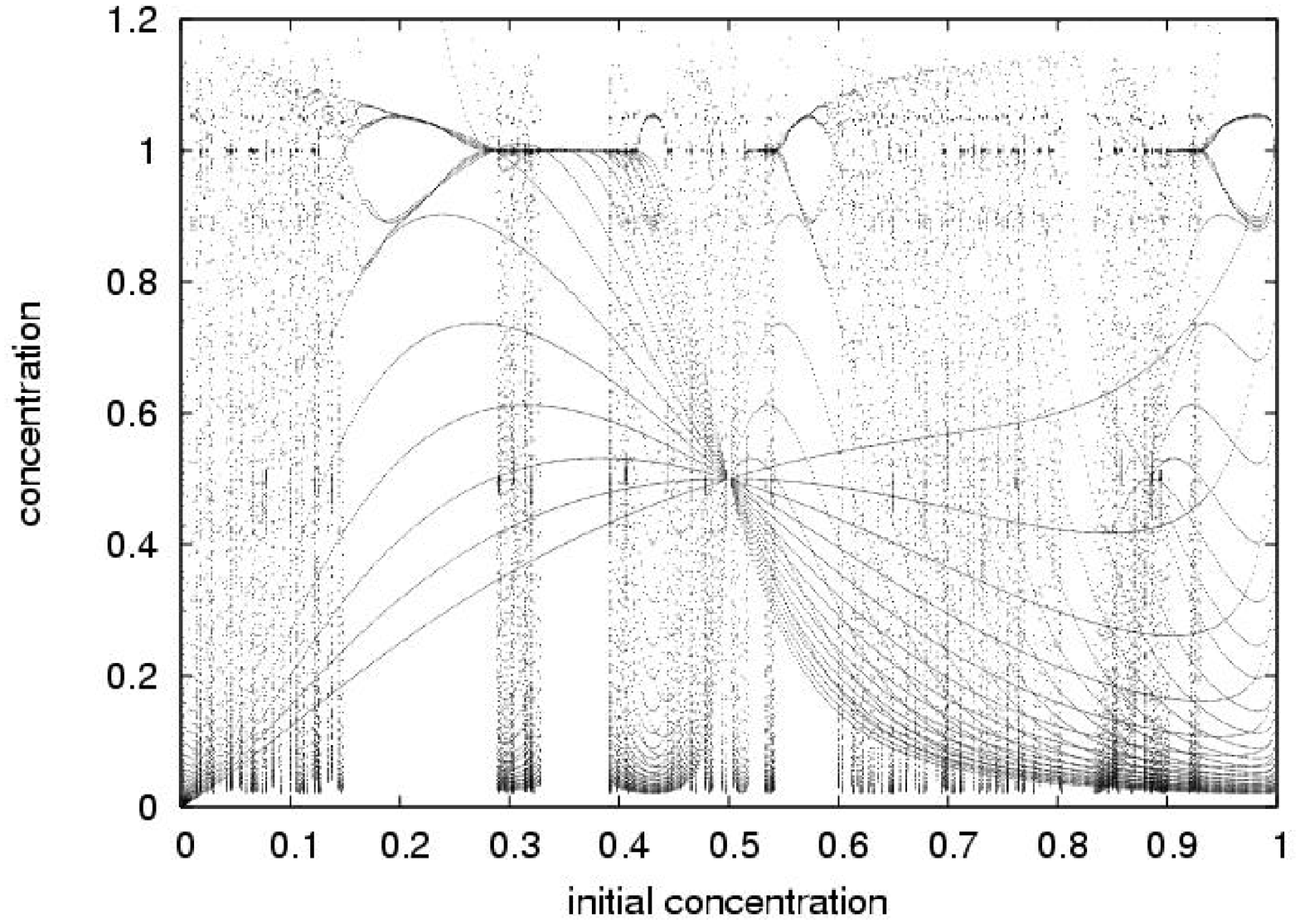}
	\caption{Coevolution of the system as a function of its initial concentration. The coevolution of a system is represented by a vertical series of dots; $t_d = 15 \, IS$}
	\label{fig:15}
\end{figure}

\begin{figure}
	\centering
	\includegraphics[angle=-90,scale = 0.45]{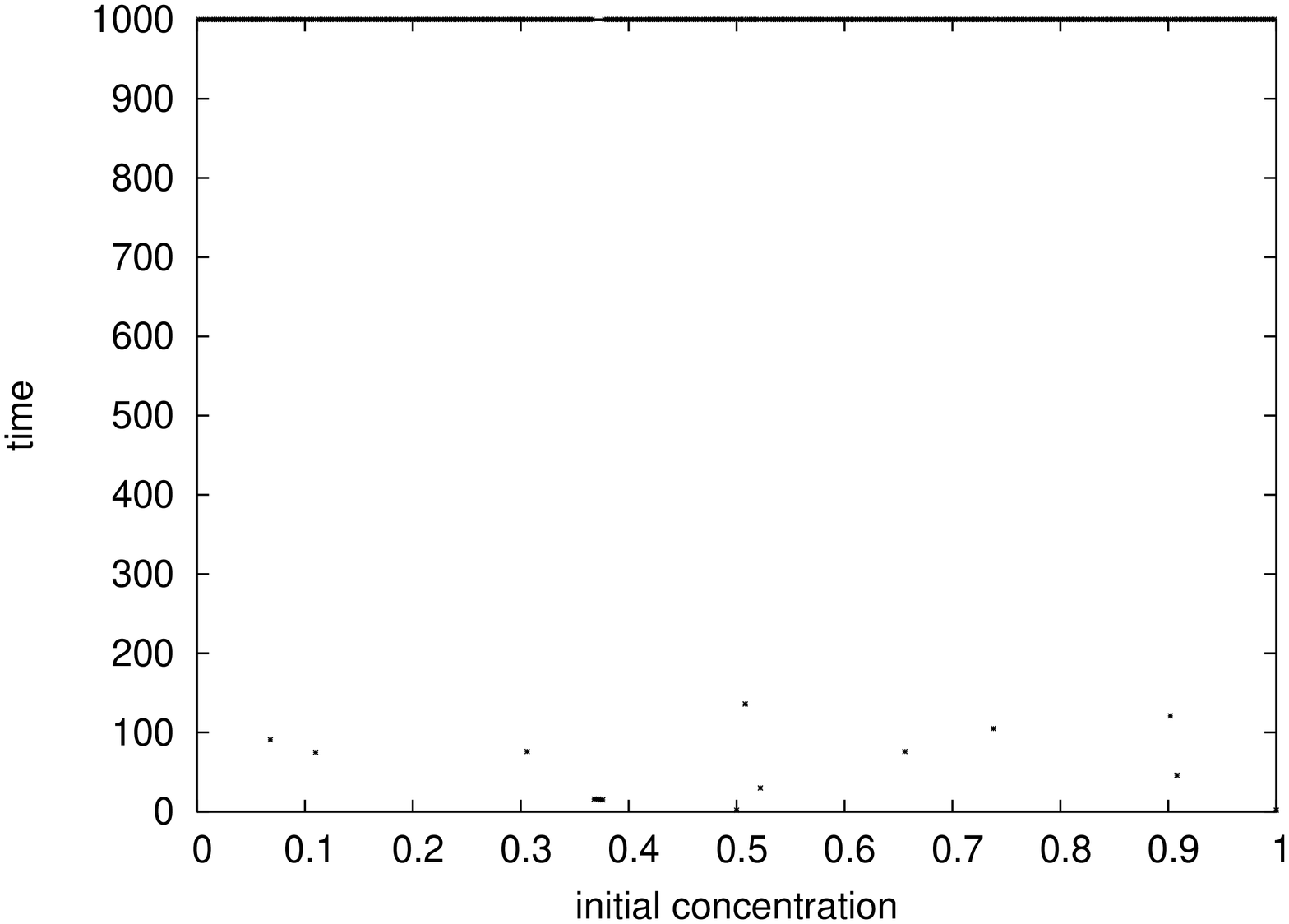}
	\caption{The crash time of the system as a function of its initial
concentration; $t_d=15 \, IS$}
	\label{fig:asymp15}
\end{figure}

Extending the delay time above six iteration steps leads to a possible
collapse of the system. For $t_d \geq 7 \, IS$ the system may crash. The crash is
defined when the concentration of companies becomes negative or zero.
Examples of such evolutions which lead to a crash are presented in the case of
$t_d =12 IS$ and $t_d=15 IS$ on  Fig. \ref{fig:12} and Fig. \ref{fig:15}
respectively. The crash of the system is presented in such plots as a white
band containing very few points in the vertical direction. It is also seen in Fig.
\ref{fig:asymp15}, where for several intervals on the initial concentration axis, e.g. $
t_d \in (0.15 ; 0.2) \cup (0.34 ; 0.36) \cup (0.53 ; 0.61)$ the crash of the
system occurs very quickly. However there are some initial concentrations for which the
evolution of the system before crash time is quite long (up to 400 IS). Additionally
in the case of $t_d = 15$, the system may evolve toward a stable state, with a
full occupation of the environment by companies.  Examples of such an evolution 
as a function of time for given initial concentrations are presented in Fig.
\ref{fig:hist12} and Fig. \ref{fig:hist15} for the cases $t_d =12$ and $t_d = 15 $
respectively.


\section{Conclusions}

Economic cycle causes and occurrences are fascinating and relevant
subjects of interest in many economy questions
\cite{Kalecki1}-\cite{others2c}.  
The
problem has been studied also by means of sociology technics \cite{Hohnisch},
showing that changes of opinions about recession  or prosperity undergo drastic
changes from one equilibrium to another, both having fluctuations in stochastically
resonant systems.
In the present investigation, an information flow, typical of economy
systems, has been incorporated into the ACP model \cite{ACP1,ACP2,ACP3}.
`This has led to observe different forms of so called cycles,
through concentration oscillations. In the case of $short$ delay time $t_d
\in (2\, IS, 4 \, IS)$, between data acquisition and policy implementation by a company, the system evolves toward a
 unique stable equilibrium state. This situation can be highly welcomed in some economy systems. Indeed this indicates that,
through an information control, a system can insure the existence of a high number of companies, whence
not threatening the system of a collapse. 

In the case of $medium$ size delay times $t_d =5 \, IS $ or $t_d =6 \, IS $,
 the system undergoes oscillations: stable concentration cycles appear in the system. This form of
evolution is often observed in economy, e.g. agricultural markets, where without
external control the level of agricultural production oscillates between over-
and underproduction. Since the enlarging of the delay time leads to the possibility
of the system to crash, such a system may require some external (governmental) control,
for its stability. In reality, the delay of information flow and policy
implementation may also fluctuate.
For long information flow delay times, $t_d \geq 7$ , the systems may crash for most 
initial concentrations. However, despite the  frequent possibility of the system
to crash the
situation is not hopeless because the crash time in many cases is long enough to
allow for some particular
control and to avoid the collapse of the company concentration. It is also
possible to observe a "economy resonance" where despite a long delay  time the
system evolves for a long time or can even reach a stable state, which insures its
existence. This latest observation is especially interesting for market
control purposes, because it points to the existence of initial conditions for which
the system may evolve during a very long time, which is vital for
the  possibility of creating
and applying some control procedures.
\index{time delay - initial concentration resonance}

\section{Acknowledgement}
JM stay in Liege was partially supported by a STSM grant funded by the COST P10 
Physics of Risk program.

\printindex

 \bibliographystyle{unsrt}
 \bibliography{verhulst.bib}

\end{document}